\documentclass[superscriptaddress,twocolumn,showpacs]{revtex4}
\usepackage{amsmath,amssymb}
\usepackage{graphicx}

\begin{document}

\title{Phase Separation of Multi-Component Bose-Einstein Condensates of Trapped Atoms and Molecules with a Homonuclear Feshbach Resonance}

\author{Ryosuke Shibato}
\affiliation{Department of Physics, Tokyo Metropolitan University,  Hachioji, Tokyo 192-0397, Japan}

\author{Takushi Nishimura}
\affiliation{Division of Advanced Sciences, Ochadai Academic Production, Ochanomizu University, Otsuka, Bunkyo, Tokyo 112-8610, Japan}

\author{Takayuki Watanabe}

\author{Toru Suzuki}
\affiliation{Department of Physics, Tokyo Metropolitan University,  Hachioji, Tokyo 192-0397, Japan}

\date{\today}

\begin{abstract}

We investigate phase separation of Bose-Einstein condensates (BECs) of two-component atoms and one-component molecules with a homonuclear Feshbach resonance. We develop a full model for dilute atomic and molecular gases including correlation of the Feshbach resonance and all kinds of interparticle interactions, and numerically calculate order parameters of the BECs in spherical harmonic oscillator traps at zero temperature with the Bogoliubov's classical field approximation. As a result, we find out that the Feshbach resonance can induce two types of phase separation. The actual phase structures and density profiles of the trapped gases are predicted in the whole parameter region, from the atom dominant regime to the molecule dominant regime. We focus on the role of the molecules in the phase separation. Especially in the atom dominant regime, the role of the molecules is described through effective interactions derived from our model. Furthermore we show that a perturbative and semi-classical limit of our model reproduces the conventional atomic BEC (single-channel) model. 

\end{abstract}

\pacs{03.75.Mn, 05.30.Jp, 67.85.Fg}

\maketitle

\renewcommand{\vec}[1]{\mbox{\boldmath $#1$}}


\section{Introduction \label{Sec-intro}}

Since the realization of Bose-Einstein condensation in alkali atoms in 1995~\cite{BEC_Rb,BEC_Na,BEC_Li}, studies on cold atomic gases have greatly advanced~\cite{PandS}. The cold atoms are precisely controlled by electromagnetism and optics, and flexible to design quantum systems. In particular, interatomic interactions are effectively controlled in the Feshbach resonance technique~\cite{PandS}, which realizes, e.g., artificial collapse of atomic BECs~\cite{BN_exp}, BEC-BCS crossover in Fermi gases~\cite{BECBCS}, and phase separation of multi-component BECs~\cite{papps,tojo} related to this work. 

In this paper, we consider BECs of multi-component atoms. The components of the atoms are determined by their atomic species, mass numbers, and spin states and manually designed in actual experiments. In fact, two-component BECs of ${}^{87}$Rb atoms with different spin states and three-component BECs of ${}^{23}$Na atoms with spin triplet (spinor BECs) are produced in the end of 1990s~\cite{bb_1_exp, bb_2_exp, bb_3_exp}. Physics of the multi-component BECs is one of the current issues in quantum physics.

The multi-component BECs can exhibit immiscible phase separation, where the BECs do not exist as a mixture in equilibrium but produce domain structures made of different components of the BECs. Although such a phase separation is commonly seen in various liquids, the phase separation of dilute gases is a special feature of quantum systems dominated by energy in stead of entropy inducing the miscibility of classical gases. In fact, the phase separation of two-component atomic BECs is due to competition of the repulsive energies between same and different components of the BECs~\cite{bb_the1, bb_the2, bb_the3}. The phase separation reflects quantum statistics and interactions and is thus important in quantum physics.

The behaviors of the atomic BECs, e.g., the phase separation, can be controlled in the Feshbach resonance technique. In this technique, molecules made of the atoms are introduced through the Feshbach resonance, which is a resonance between an interatomic scattering state (open-channel) and a molecular bound state with a different spin structure (closed-channel). As a result, the molecules work as a resonant intermediary and effectively modifies the interaction in the open-channel. Energy detuning of the resonance can be controlled with external magnetic field owing to the Zeeman effect~\cite{hetero1}. Recently two experimental groups have succeeded in inducing the phase separation in this technique: JILA group demonstrates it in BECs of ${}^{85}$Rb and ${}^{87}$Rb atoms with homonuclear ${}^{85}$Rb molecules~\cite{papps}; Gakushuin group demonstrates it in the spinor BECs of ${}^{87}$Rb atoms~\cite{tojo}.

The purpose of this paper is to describe the phase separation and domain structures of the multi-component BECs of the atoms and molecules with the homonuclear Feshbach resonance. Here the molecules can play an active role and we are particularly interested in the relationship between the molecules and phase structures.

In order to accomplish the purpose, we develop a full model for the atoms and molecules including correlation of the Feshbach resonance, all kinds of interparticle interactions, and trap potentials. This model enables to predict the actual phase structures and density profiles of the trapped BECs in the whole parameter region, from the atom dominant regime to the molecule dominant regime, and to investigate the overall behaviors of the molecules. 

In general, molecular formations in quantum gases can induce various phase structures. In quantum many-particle physics, molecules appear as a kind of many-body correlations, e.g., the molecule-pair (BEC-BCS) crossover for the molecule made of a pair of fermions~\cite{NSL}. The crossover theory is originally developed in the fermionic systems and adapted in the boson-fermion environments~\cite{bf_pair}, four-body correlations~\cite{sogo}, and so on. In particular, one of the authors and his collaborators develop quasi-chemical equilibrium theory for diatomic molecules to the all kinds of quantum statistical environments including the bosonic environment related to the present work~\cite{TN}. In addition, the molecular formations in the bosonic systems are realized in actual experiments~\cite{FM_1, FM_2} and studied in other theories~\cite{fm_1, fm_2}. 

In the atom dominant regime, the contribution of the molecules can be described as effective interactions for a partial atomic space in the full model, where the residual molecular space induces the effective interactions owing to the atom-molecule interactions and correlation. In general, the partial system and effective interactions are artificially defined from the full system and not unique. The theory of the effective interactions is a powerful tool to study physics~\cite{Brandow}. 

A conventional approach to treat the Feshbach resonance in the atom dominant regime is introduction of an effective interaction to the atomic BEC (single-channel) model in a perturbative way~\cite{PandS, BN_the1, BN_the2}, where the molecules implicitly contribute to the atoms only through the effective interaction. In Sec~\ref{Sec-discus}, we show that a perturbative and semi-classical limit of our model reproduces the single-channel model. 

This paper is organized as follows. In section~\ref{Sec-form}, we propose our model and derive formulas with some approximations. In section~\ref{Sec-result}, we show a macroscopic analysis of the phase separation and numerical results of the phase structures and density profiles of the trapped gases. In section~\ref{Sec-discus}, we define the effective interactions in our model to see the role of the molecules in the atom dominant regime. In addition, we discuss the correspondence relation and difference between our model and the single-channel model. Section~\ref{Sec-summary} is devoted to summary and perspective.

%
%
%
%
%
%
\section{Formulation \label{Sec-form}}

We consider the quantum gases of two-component bosonic atoms, labeled $a$ and $b$, and one-component molecules, labeled $m$, formed from pairs of the $a$ atoms. Assume that the gases are trapped in vacuum, and total numbers, $N_{a t}$ and $N_{b t}$, of the $a$ and $b$ atoms are fixed and related to respective particle numbers, $N_{a}$, $N_{b}$, and $N_{m}$, of the $a$ atoms, $b$ atoms, and molecules as 
\begin{equation}
N_{a t} 
= N_{a} + 2 N_{m}, 
~~
N_{b t} 
= N_{b}, 
\label{Eq-form-1}
\end{equation}
where $N_{a}$ and $N_{m}$ can vary through the Feshbach resonance. 

In this paper, the subscripts for the labels $a$, $b$, and $m$ indicate the corresponding particles. In addition, we take $\hbar = 1$ for the reduced Planck constant as a convention. 

The Hamiltonian is denoted by
\begin{equation}
H 
= \left[ \sum_{\alpha = a, b, m} H_{\alpha} \right] + \left[ \sum_{\alpha = a, b, m} \sum_{\beta = a, b, m} H_{\alpha \beta} \right] + H_{\text{FR}} 
\label{Eq-form-2}
\end{equation}
with the free-motion parts, $H_{\alpha}$, interparticle interaction parts, $H_{\alpha \beta}$, and Feshbach resonance part, $H_{\text{FR}}$. 

The free-motion part $H_{\alpha}$ in eq.~(\ref{Eq-form-2}) is defined as 
\begin{equation}
H_{\alpha} 
= \int d{\vec{r}}~ \Psi_{\alpha}^{\dagger}(\vec{r}) h_{\alpha}(\vec{r}) \Psi_{\alpha}(\vec{r}) 
\label{Eq-form-3}
\end{equation}
with the boson field operator $\Psi_{\alpha}(\vec{r})$ obeying the commutation relations, 
\begin{equation}
\big[ \Psi_{\alpha}(\vec{r}), \Psi_{\beta}(\vec{r}^{\prime}) \big] 
= 0 
\label{Eq-form-4}
\end{equation}
and 
\begin{equation}
\big[ \Psi_{\alpha}(\vec{r}), \Psi_{\beta}^{\dagger}(\vec{r}^{\prime}) \big] 
= \delta_{\alpha \beta} \delta(\vec{r} - \vec{r}^{\prime}). 
\label{Eq-form-5}
\end{equation}
The single-particle Hamiltonian $h_{\alpha}(\vec{r})$ in eq.~(\ref{Eq-form-3}) is given as 
\begin{equation}
h_{\alpha}(\vec{r}) 
= - \frac{\nabla^{2}}{2 m_{\alpha}} + V_{\alpha}(\vec{r}) + M_{\alpha} 
\label{Eq-form-6}
\end{equation}
with the particle mass $m_{\alpha}$, trap potential $V_{\alpha}(\vec{r})$, and mass term $M_{\alpha}$. In the nonrelativistic model, the values of the mass terms have no meaning except for the relative value between those of the atom and molecule. Thus we can put $M_{a} = M_{b} = 0$ and $M_{m} = \Delta{E}(\vec{r}) + 2 V_{a}(\vec{r}) - V_{m}(\vec{r})$ with the energy detuning 
\begin{equation}
\Delta{E}(\vec{r}) 
\equiv V_{m}(\vec{r}) + M_{m} - 2 (V_{a}(\vec{r}) + M_{a}) 
\label{Eq-form-6-2}
\end{equation}
without loss of generality. The energy detuning $\Delta{E}(\vec{r})$ can be varied by controlling the external magnetic field owing to the Zeeman effect. 

The interaction part $H_{\alpha \beta}$ in eq.~(\ref{Eq-form-2}) is defined as 
\begin{equation}
H_{\alpha \beta} 
= \frac{g_{\alpha \beta}}{2} \int d{\vec{r}}~ \Psi_{\alpha}^{\dagger}(\vec{r}) \Psi_{\beta}^{\dagger}(\vec{r}) \Psi_{\beta}(\vec{r}) \Psi_{\alpha}(\vec{r}) 
\label{Eq-form-7}
\end{equation}
with the contact-type interactions in the pseudo-potential approach. This approach must be valid for the low-energy scattering dominated by the $s$-wave scattering length and also for the interactions in the cold dilute gases. For the BECs, we can put 
\begin{equation}
g_{\alpha \beta} 
= g_{\beta \alpha} 
= 2 \pi \frac{m_{\alpha} + m_{\beta}}{m_{\alpha} m_{\beta}} a_{\alpha \beta} 
\label{Eq-form-8}
\end{equation}
with the bare $s$-wave scattering length $a_{\alpha \beta}$ between the $\alpha$ and $\beta$ particles in vacuum according to the regularization scheme. This approach has been widely applied to the cold atoms and molecules~\cite{PandS}. Note that the interaction part in eq.~(\ref{Eq-form-7}) does not include effect of the Feshbach resonance correlation. 

We here supplement explanation of the regularization procedure for the contact-type interactions. In general, the coupling constants, $g_{\alpha \beta}$, are theoretical dummy parameters to reproduce relations among physical values through the regularization procedure. In fact, direct perturbative calculation for $g_{\alpha \beta}$ can produce the divergence difficulty owing to insufficient treatment of singularity of the two-body wave-functions. In the above paragraph, to deal with the difficulty, we introduce the renormalized coupling constants in eq.~(\ref{Eq-form-8}) with the two-body wave-functions in vacuum. That is because the singularity of the two-body wave-functions is assumed as a extremely-short-range property separated from the global property described in the asymptotic fields for the BECs. 

The Feshbach resonance part $H_{\text{FR}}$ in eq.~(\ref{Eq-form-2}) is defined as
\begin{equation}
H_{\text{FR}} 
= \lambda \int d{\vec{r}} \left[ \Psi_{m}^{\dagger}(\vec{r}) \big( \Psi_{a}(\vec{r}) \big)^{2} + \big( \Psi_{a}^{\dagger}(\vec{r}) \big)^{2} \Psi_{m}(\vec{r}) \right] 
\label{Eq-form-9}
\end{equation}
also in the pseudo-potential approach, where we apply the same regularization procedure for the BECs. The absolute value of the $\lambda$ in eq.~(\ref{Eq-form-9}) can be obtained from the detuning dependence of the $s$-wave scattering length with the Feshbach resonance in vacuum, 
\begin{equation}
\tilde{a}_{a a}(\Delta{E}) 
\approx a_{a a} - \frac{m_{a}}{2 \pi} \frac{\lambda^{2}}{\Delta{E}}, 
\label{Eq-form-10}
\end{equation}
as shown in appendix~\ref{App-fr}. In addition, we can freely determine the sign of the $\lambda$ in eq.~(\ref{Eq-form-9}) because it is related to the phases of the field operators. The $\tilde{a}_{a a}(\Delta{E})$ and $a_{a a}$ in eq.~(\ref{Eq-form-10}) can be obtained in scattering experiments in principle. No free parameter thus survives in this model. 

Here we apply the Bogoliubov's classical field approximation~\cite{BCFA}. First we introduce BECs defined with the BEC order parameters, $\phi_{\alpha}(\vec{r}) \equiv \big< \Psi_{\alpha}(\vec{r}) \big>$, by using the Bogoliubov's method, $\Psi_{\alpha}(\vec{r}) = \phi_{\alpha}(\vec{r}) + \tilde{\Psi}_{\alpha}(\vec{r})$. Next we neglect the thermal and correlational part, 
\begin{equation}
\Psi_{\alpha}(\vec{r}) 
\approx \phi_{\alpha}(\vec{r}), 
\label{Eq-form-11}
\end{equation}
in the zero-temperature equilibrium. This approximation must be valid in BEC dominant regime as shown in the actual experiments~\cite{papps,tojo}, and can not be applied to strongly-correlated Bose gases with little or no BECs. The latter situation is beyond the purpose of this paper. 

The order parameters, $\phi_{\alpha}(\vec{r})$, in the zero-temperature equilibrium are obtained from minimization of the energy given in eq.~(\ref{Eq-form-2}) with eq.~(\ref{Eq-form-11}) under the number constraint in eq.~(\ref{Eq-form-1}) with 
\begin{equation}
N_{\alpha} 
\approx \int d{\vec{r}} \big| \phi_{\alpha}(\vec{r}) \big|^{2}. 
\label{Eq-form-12}
\end{equation}
In the Lagrange's multiplier method, we obtain the coupled non-linear Schr\"{o}dinger-type equations, 
\begin{equation}
\mu_{a} \phi_{a} 
= \bigg( h_{a}(\vec{r}) + \sum_{\alpha = a, b, m} g_{a \alpha} \big| \phi_{\alpha} \big|^{2} \bigg) \phi_{a} + 2 \lambda \phi_{m} \phi_{a}^{*}, 
\label{Eq-form-13}
\end{equation}
\begin{equation}
\mu_{b} \phi_{b} 
= \bigg( h_{b}(\vec{r}) + \sum_{\alpha = a, b, m} g_{b \alpha} \big| \phi_{\alpha} \big|^{2} \bigg) \phi_{b}, 
\label{Eq-form-14}
\end{equation}
and 
\begin{equation}
\mu_{m} \phi_{m} 
= \bigg( h_{m}(\vec{r}) + \sum_{\alpha = a, b, m} g_{m \alpha} \big| \phi_{\alpha} \big|^{2} \bigg) \phi_{m} + \lambda \phi_{a}^{2}, 
\label{Eq-form-15}
\end{equation}
and chemical equilibrium equation, 
\begin{equation}
2 \mu_{a} 
= \mu_{m}, 
\label{Eq-form-16}
\end{equation}
for the chemical potentials, $\mu_{\alpha}$. 

Lastly we summarize the treatment of BEC in the present approach. In principle, BEC is a phase transition defined in macroscopic systems; On another front, the confinement traps produce microscopic systems, where the phase transition is not well-defined owing to the nonnegligible finite size effect. In order to introduce BEC to the microscopic system, we must determine the treatment of the finite size effect. In the present approach, we firstly apply the Bogoliubov's method in eq.~(\ref{Eq-form-11}) to introduce BEC without the finite size effect, and afterward introduce the traps and finite size effect. As a result, the BECs are defined in the microscopic system with the finite size effect. In fact, the phase separation of the BECs appears not as a sharp density separation but as a smooth density crossover as demonstrated in the next section ~\footnote{%
The present approach is merely a theory to give an effective model for the microscopic system. As another approach, we can apply the variational method with a trial state denoted by the coherent state for the bosons. Then we obtain the same formulation after all. 
}.

\section{Phase separation and domain structures \label{Sec-result}}

In this section, we show results of the calculation according to eqs.~(\ref{Eq-form-12})-(\ref{Eq-form-16}). Parameters are defined in subsection~\ref{Sec-result-para}. Calculational scheme is given in subsection~\ref{Sec-result-cal}. We show general property and typical behaviors of the density profiles in subsection~\ref{Sec-result-gene} and phase structures in subsection~\ref{Sec-result-phase}. 
\subsection{Parameters \label{Sec-result-para}}

In order to demonstrate the calculation, we assume a situation as follows: First the $a$ and $b$ atoms have mostly the same masses, $m_{a} = m_{b} \equiv m$, and coupling constants, 
\begin{equation}
g_{a a} 
= g_{a b} 
= g_{b b} 
\equiv g. 
\label{Eq-result-0}
\end{equation}
Second the atoms and molecules are trapped in the spherical harmonic oscillator potentials denoted by 
\begin{equation}
V_{\alpha}(\vec{r}) 
= \frac{1}{2} m_{\alpha} \omega^{2} r^{2} 
\label{Eq-result-1}
\end{equation}
with the trap frequency $\omega$. Here $m_{m} \approx 2 m$ for the nonrelativistic diatomic molecule, and the spatial dependence of the energy detuning $\Delta{E}(\vec{r})$ in eq.~(\ref{Eq-form-6-2}) is negligible. This assumption corresponds to the case of atoms with two spin-components, and leads to a typical demonstration of the phase separation. 

We also assume the hard-sphere hypothesis: The atoms and molecule have hard-sphere with radii $R_{a}$ ($= R_{b}$) and $R_{m}$, respectively, and the scattering lengths in eq.~(\ref{Eq-form-8}) become 
\begin{equation}
a_{\alpha \beta} 
= R_{\alpha} + R_{\beta}. 
\label{Eq-result-2}
\end{equation}
According to eq.~(\ref{Eq-form-8}), we then obtain 
\begin{equation}
g_{a m} 
= g_{b m} 
= \frac{3}{8} \left( 1 + \kappa \right) g 
\label{Eq-result-3}
\end{equation}
and 
\begin{equation}
g_{m m} 
= \frac{\kappa}{2} g 
\label{Eq-result-4}
\end{equation}
with $\kappa \equiv R_{m} / R_{a}$, where $g_{\alpha \beta}$ can be described only with the two positive parameters $g$ and $\kappa$. Note that this hypothesis is just for convenience to fix the parameters and simplify the model for demonstration. In principle, the scattering lengths should be determined in scattering experiments. 

We summarize the parameters. In the harmonic oscillator units, $m = 1$, $\omega = 1$, and $\hbar = 1$ without loss of generality, and the remaining parameters are $g$, $\kappa$, $\lambda$, $N_{b}$, $N_{a t}$ ($= N_{a} + 2 N_{m}$), and $\Delta{E}$. Note that the number ratio of the molecules, $N_{m} / N_{a t}$ , is a monotonically decreasing function of the detuning energy, $\Delta{E}$, and we can then take $N_{m}$ as a parameter instead of $\Delta{E}$. 


\subsection{Calculational scheme \label{Sec-result-cal}}

In order to compute eqs.~(\ref{Eq-form-13})-(\ref{Eq-form-15}), we utilize the imaginary-time relaxation method with the spatial difference scheme. In this method, we successively improve a trial state starting from an input initial state to obtain a solution. Thus the resulting solution may generally depend on the initial state if several solutions coexist in a numerical precision. 

In this paper, we give the initial state in the following scheme. First we solve them with an arbitrary initial state at $N_{m} \sim 0$, where we can not find the initial state dependence. Next we successively solve them with asymptotical increment of $N_{m}$ by substituting the obtained state with tiny numerical noises to avoid numerical instability for the next initial state. We iterate this process until $N_{m}$ reaches a given number. 

In addition, we also perform the same scheme but started from $N_{m} \sim N_{a t} / 2$ with asymptotical decrement of $N_{m}$. As a result, we confirm that both results agree in the numerical precision. 


\subsection{General property \label{Sec-result-gene}}

The phase separation of the BECs is due to a discontinuous phase transition in terms of a ratio between two of the component densities. This phase transition can be described as instability for local density fluctuations in the uniform mixture. In general, the instability indicates not only the phase separation -- appearance of lower limits of the densities in equilibrium -- but also collapse -- appearance of the divergent densities. If a system does not induce the collapse, the instability is directly linked to the phase separation. 

In the present system, the instability indicates the phase separation because this system has no attraction to induce the collapse. In fact, the only attraction in this system is the Feshbach resonance part in eq.~(\ref{Eq-form-9}), which can not induce the collapse because the power of the density dependence of the attraction energy in eq.~(\ref{Eq-form-9}), $3 / 2$, is less than that of the repulsion energy in eq.~(\ref{Eq-form-7}), $2$.

The stability condition at the zero temperature can be described as minimum condition of the energy density functional with the number constraints. The detail of the stability condition is summarized in appendix~\ref{App-psc}. 

The stability condition suggests that the three-component mixture is always unstable except for $\lambda = 0$ and breaks into immiscible domains. In other words, this system exhibits the phase separation whenever the Feshbach resonance correlation in eq.~(\ref{Eq-form-9}) exists. This result depends on the assumption for the atoms and hard-sphere hypothesis in eqs.~(\ref{Eq-result-3}) and (\ref{Eq-result-4}), and is independent of values of the parameters in subsection~\ref{Sec-result-para}. 

Moreover, according to the stability condition, possible phases in the immiscible domains are limited to ($b$), ($m$), and ($a$, $m$), where the labels in the round brackets represent the components of the existing particles in the equilibrium. The phase ($b$) is always stable or metastable while the other phases have their own stability conditions described in appendix~\ref{App-psc}. In particular, the phase ($m$) has a lower limit for the molecular density $n_{m}$ and is stable or metastable for the larger $n_{m}$ than the limit. 

In the microscopic trap system, the phase separation does not make the sharp domain structure except for the theoretical Thomas-Fermi limit, $N_{a t}, N_{b} \to \infty$, because of the finite size effect. In the actual experiments, the numbers of the atoms are limited below $10^{4 \sim 6}$ in the evaporative cooling method, and ranges of interfaces between the domains, i.e., the healing lengths, are not negligible. 

Furthermore the above macroscopic analysis for the phase separation does not give any information for profiles and compositions of the domains. That is because they depend on the interface effects, which can not be given as extensive thermodynamical variables. In order to obtain the actual domain structure, the microscopic analysis is needed beyond the macroscopic analysis. 

\begin{figure}[ht]
\begin{center}
\includegraphics[width=7cm]{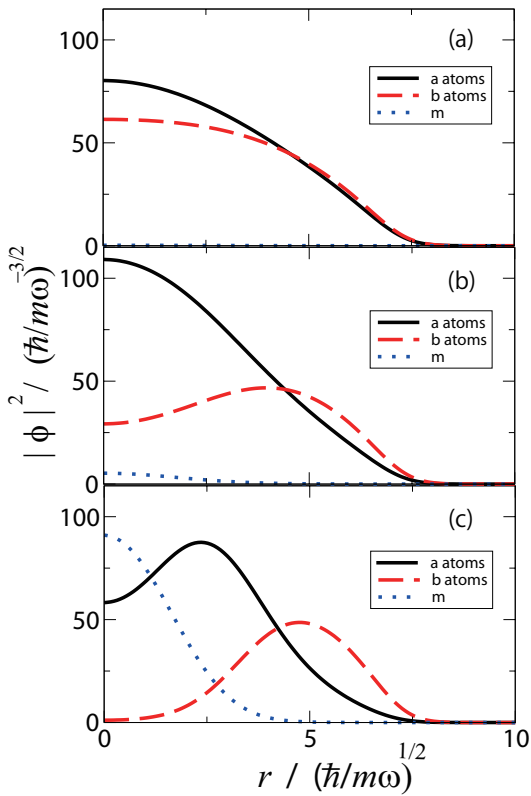}
\end{center}
\caption{(Color online) 
The density profiles when $g = 0.1 \times m^{- 3 / 2} \omega^{- 1 / 2}$, $\kappa = 2$, $\lambda = 0.1 \times m^{- 3 / 4} \omega^{1 / 4}$, $N_{a t} = N_{b} = 5 \times 10^{4}$, and $N_{m} = 5 \times 10^{1}$ (a), $5 \times 10^{2}$ (b), and $5 \times 10^{3}$ (c). The solid, dashed, and dotted lines denote those of the $a$ atoms, $b$ atoms, and molecules, respectively. 
}
\label{den}
\end{figure}

In fig.~\ref{den}, we show typical behaviors of the density profiles in the spherical system with $r \equiv \big| \vec{r} \big|$ when $g = 0.1 \times m^{- 3 / 2} \omega^{- 1 / 2}$, $\kappa = 2$, $\lambda = 0.1 \times m^{- 3 / 4} \omega^{1 / 4}$, $N_{a t} = N_{b} = 5 \times 10^{4}$, and $N_{m} = 5 \times 10^{1}$ (a), $5 \times 10^{2}$ (b), and $5 \times 10^{3}$ (c). If we change the values of the parameters as in the next subsection, the qualitative behaviors are mostly similar to those in fig.~\ref{den}. 

Fig.~\ref{den}a represents the profiles in the atom dominant regime, $N_{m} / N_{a t} = 10^{-3}$. Although $N_{m} / N_{a t} \approx 0$, the molecules affect the atomic distributions, which are obviously different from the symmetric ones, $\big| \phi_{a}(\vec{r}) \big|^{2} = \big| \phi_{b}(\vec{r}) \big|^{2}$, at the $N_{m} = 0$ limit. It is due to the Feshbach resonance part in eq.~(\ref{Eq-form-9}) and suggests the phase separation.

The phase separation is more clearly seen in fig.~\ref{den}b for $N_{m} / N_{a t} = 10^{-2}$. As $N_{m}$ increases, the $b$ atoms are gradually localized outside the $a$ atoms and molecules. According to the macroscopic analysis, these profiles indicate an unsharp double-domain structure made of the two phases ($a$, $m$) and ($b$) in common with those in fig.~\ref{den}a. 

Fig.~\ref{den}c represents the density distributions in the atom-molecule coexistence regime, $N_{m} / N_{a t} = 10^{-1}$. As $N_{m}$ increases, the $a$ atoms are gradually localized in the middle region between the inner molecules and outer $b$ atoms. According to the macroscopic analysis, these profiles indicate an unsharp triple-domain structure made of the three phases ($m$), ($a$, $m$), and ($b$). Here we emphasize that this structure can not be seen in the conventional single-channel model, which does not include the molecular phase ($m$) and must be valid only in the atom dominant regime. 

The constitution of the unsharp domain structures is due to minimization of the energy made up of the kinetic and trap parts in eq.~(\ref{Eq-form-3}), interaction parts in eq.~(\ref{Eq-form-7}), and Feshbach resonance part in eq.~(\ref{Eq-form-9}). The domain profiles principally reflect the surface and interface energies derived from the kinetic parts. The domain arrangement principally reflects the energy competition among the trap, interaction, and Feshbach resonance parts. In a rough explanation, the $a$ atoms tend to take the center position in the atom dominant regime (figs.~\ref{den}a and b) because of the attraction in the Feshbach resonance part, which gives an advantage to enlarge the densities of the $a$ atoms and molecules; the molecules tend to take the center position in the atom-molecule coexistence regime (fig.~\ref{den}c) because of the mass difference in the trap potentials in eq.~(\ref{Eq-result-1}) and oscillator lengths, $1 / \sqrt{m_{\alpha} \omega}$.

\subsection{Phase structures \label{Sec-result-phase}}

As demonstrated above, the phase separation induces the unsharp domain structures in the microscopic system. Some of the density profiles can take a hollow structure in the spherical system as shown in fig.~\ref{den}. 

In order to describe phase structures of the density profiles, we introduce indicator quantities defined as 
\begin{equation}
\upsilon_{\alpha} 
= \frac{\big| \phi_{\alpha}(0) \big|^2}{\max{\left[ \big| \phi_{\alpha}(\vec{r}) \big|^2 \right]}}
\label{Eq-result-phase-1}
\end{equation}
for $\alpha = a$, $b$, and $m$. For the non-hollow-type profile, the maximal density is at the center, $\vec{r} = 0$, and $\upsilon_{\alpha} = 1$; otherwise $\upsilon_{\alpha} < 1$. 

\begin{figure}[ht]
\begin{center}
\includegraphics[width=7cm]{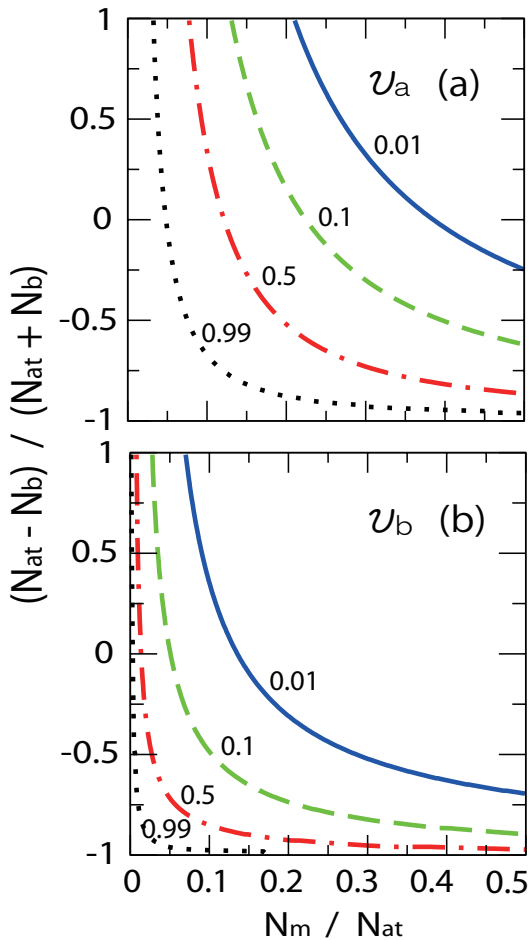}
\end{center}
\caption{(Color online) 
The phase structures of the density profiles of the $a$ atoms (a) and $b$ atoms (b) when $g = 0.1 \times m^{- 3 / 2} \omega^{- 1 / 2}$, $\kappa = 2$, $\lambda = 0.1 \times m^{- 3 / 4} \omega^{1 / 4}$, and $N_{a t} + N_{b} = 1.0 \times 10^5$. The solid, dashed, dot-dashed, and dotted lines denote the contour lines for $\upsilon_{\alpha} = 0.01$, $0.1$, $0.5$, and $0.99$, respectively. 
}
\label{contourline}
\end{figure}

In fig.~\ref{contourline}, we plot the phase structures for the $a$ atoms (a) and $b$ atoms (b) with the contour lines for the $\upsilon_{\alpha}$ in eq.~(\ref{Eq-result-phase-1}), where we omit to plot for $\upsilon_{m}$ because the molecules do not exhibit the hollow-type profile and $\upsilon_{m} = 1$. Here we choose the same values in fig.~\ref{den} for $g$, $\kappa$, and $\lambda$ and give the total number of the atoms as $N_{a t} + N_{b} = 10^5$. The cases in fig.~\ref{den} correspond to those on the $N_{a t} - N_{b} = 0$ line in fig.~\ref{contourline}. Note that edges of the variable area in fig.~\ref{contourline} are singular, where one of the particle numbers is zero. 

Fig.~\ref{contourline} represents the smoothly-varying phase structures. In the smaller $N_{m}$ regions than the $\upsilon_{\alpha} = 0.99$ lines, $\upsilon_{\alpha} \approx 1$ and the corresponding particles take the non-hollow-type profile. In the larger $N_{m}$ regions than the $\upsilon_{\alpha} = 10^{-2}$ lines, $\upsilon_{\alpha} \approx 0$ and the corresponding particles completely take the hollow-type profile. The value of $\upsilon_{a}$ is always larger than that of $\upsilon_{b}$. It indicates that, as $N_{m}$ increases, the double-domain structure appears first and afterward crosses over into the triple-domain structure. As $N_{b}$ increases, the effect of the molecules and $a$ atoms on the $b$ atoms relatively decreases, and the densities of the molecules and $a$ atoms also decrease; Then the appearance of the domain structures delays as a result. 

Lastly we comment on the dependence on the parameters fixed in fig.~\ref{contourline}. First, as $N_{a t} + N_{b}$ and $g$ increase, the system approaches to the Thomas-Fermi limit, where the domain structures appear sharply. Second, as $\big| \lambda \big|$ increases, the $a$ atoms and molecules tend to mix and concentrate owing to the attraction in the Feshbach resonance part, and the domain structures can easily appear owing to the concentration of the $a$ atoms and molecules according to stability conditions in eqs.~(\ref{Eq-psc-7}) and (\ref{Eq-psc-8}). Third $\kappa$ does not affect the qualitative properties of the phase structures as confirmed in the stability conditions and also in the numerical analysis for various values of $\kappa$.

\section{Effective interactions \label{Sec-discus}}

In this section, we discuss the role of the molecules in the phase separation and double-domain structures in the atom dominant regime by introducing the effective interactions. In addition, we show that a perturbative and semi-classical limit of the present approach reproduces the single-channel model. Validity of the single-channel model is also discussed with numerical demonstration. 

In subsection~\ref{Sec-discus-nonper}, we define the effective interactions for the present model. In subsection~\ref{Sec-discus-single}, we take the limit and obtain the single-channel model. In subsection~\ref{Sec-discus-role}, we discuss the role of the molecules. In subsection~\ref{Sec-discus-compar}, we compare the present and single-channel models.

\subsection{Exact formulation \label{Sec-discus-nonper}}

In order to define the effective interactions, we focus on a partial system only consisting of the BECs of the $a$ and $b$ atoms, where the number of the $a$ atoms, $N_{a}$, in the partial system is counted without the atoms in the molecules and does not agree with the total number in the full system, $N_{a t} \ne N_{a}$. 

Since the partial system does not include the molecules, the BEC order parameters should obey 
\begin{equation}
\tilde{\mu}_{a}(\vec{r}) \phi_{a} 
= \bigg( h_{a}(\vec{r}) + \sum_{\alpha = a, b} \tilde{g}_{a \alpha}(\vec{r}) \big| \phi_{\alpha} \big|^{2} \bigg) \phi_{a} 
\label{Eq-discus-nonper-1}
\end{equation}
and 
\begin{equation}
\tilde{\mu}_{b}(\vec{r}) \phi_{b} 
= \bigg( h_{b}(\vec{r}) + \sum_{\alpha = a, b} \tilde{g}_{b \alpha}(\vec{r}) \big| \phi_{\alpha} \big|^{2} \bigg) \phi_{b} 
\label{Eq-discus-nonper-2}
\end{equation}
in the same manner in section~\ref{Sec-form}, where the effective chemical potentials, $\tilde{\mu}_{\alpha}(\vec{r})$, and effective coupling constants, $\tilde{g}_{\alpha \beta}(\vec{r})$, can have the spatial dependence in general. 

Exact definitions of the effective quantities must not affect physical values, e.g., the order parameters and particle numbers, because physics must not depend on the artificial definitions. 

In order to obtain the exact definitions, we utilize the solutions of the full model in self-consistent framework, where the effective interactions can reproduce the original solutions obtained from eqs.~(\ref{Eq-form-12})-(\ref{Eq-form-16}). 

The effective chemical potentials, $\tilde{\mu}_{\alpha}(\vec{r})$, in eqs.~(\ref{Eq-discus-nonper-1}) and (\ref{Eq-discus-nonper-2}) are defined as 
\begin{equation}
\tilde{\mu}_{\alpha}(\vec{r}) 
\equiv \mu_{\alpha} - g_{\alpha m} \big| \phi_{m}(\vec{r}) \big|^{2} 
\label{Eq-discus-3}
\end{equation}
with the order parameter $\phi_{m}(\vec{r})$ calculated from the full model, where $\tilde{\mu}_{\alpha}(\vec{r})$ reflects the background energy derived from the atom-molecule interactions and indicates renormalization of the mass-energy owing to the effective interactions. 

The effective coupling constants, $\tilde{g}_{\alpha \beta}(\vec{r})$, in eqs.~(\ref{Eq-discus-nonper-1}) and (\ref{Eq-discus-nonper-2}) are defined as $\tilde{g}_{\alpha \beta}(\vec{r}) \equiv g_{\alpha \beta}$ except for 
\begin{equation}
\tilde{g}_{a a}(\vec{r}) 
\equiv \frac{4 \pi a_{a a}}{m_{a}} + \lambda \left[ \frac{\phi_{m}(\vec{r})}{( \phi_{a}(\vec{r}) )^{2}} + \frac{\phi_{m}^{*}(\vec{r})}{( \phi_{a}^{*}(\vec{r}) )^{2}} \right] 
\label{Eq-discus-2}
\end{equation}
with the order parameters $\phi_{a}(\vec{r})$ and $\phi_{m}(\vec{r})$ calculated from the full model, where $\tilde{g}_{a a}(\vec{r})$ is renormalized with the Feshbach resonance correlation and produces one of the effective interactions. According to eq.~(\ref{Eq-form-15}), we rewrite eq.~(\ref{Eq-discus-2}) as 
\begin{equation}
\tilde{g}_{a a}(\vec{r}) 
= \frac{4 \pi a_{a a}}{m_{a}} + \frac{2 \lambda^{2}}{\mu_{m}^{\prime}(\vec{r})} 
\label{Eq-discus-4}
\end{equation}
with 
\begin{equation}
\mu_{m}^{\prime}(\vec{r}) 
\equiv \mu_{m} - \frac{( h_{m}(\vec{r}) \phi_{m}(\vec{r}) )}{\phi_{m}(\vec{r})} - \sum_{\alpha} \tilde{g}_{m \alpha} \big| \phi_{\alpha}(\vec{r}) \big|^{2}. 
\label{Eq-discus-5}
\end{equation}
Furthermore, according to eqs.~(\ref{Eq-form-16}) and (\ref{Eq-form-13}), the $\mu_{m}$ in eq.~(\ref{Eq-discus-5}) can be rewritten as 
\begin{equation}
\mu_{m} 
= 2 \left[ \frac{( h_{a}(\vec{r}) \phi_{a}(\vec{r}) )}{\phi_{a}(\vec{r})} + \sum_{\alpha} \tilde{g}_{a \alpha} \big| \phi_{\alpha} \big|^{2} \right]. 
\label{Eq-discus-6}
\end{equation}

The exact definitions are applicable to the whole parameter region and reproduce the phase structures in the previous section. In fact, the double-domain structure is principally derived from $\tilde{g}_{a a}(\vec{r})$; The triple-domain structure reflects all of the effective quantities. 

The effective interactions provide a physical interpretation from the viewpoint of the partial system when we obtain the actual values of the effective quantities, which are determined from the full system, not the partial system, except for some trivial situations, e.g., the decoupling or perturbative limit. 
\subsection{The single-channel model \label{Sec-discus-single}}
We here take the perturbative and semi-classical limit of the present approach in the atom dominant regime, $N_{a t} \approx N_{a}$ and $N_{m} / N_{a t} \approx 0$, where the perturbation is started from vacuum of the atoms and molecule. 

In the limit, we directly obtain $N_{a t} \sim N_{a}$ and $\tilde{\mu}_{\alpha}(\vec{r}) \sim \mu_{\alpha}$ in eq.~(\ref{Eq-discus-3}) because $\phi_{m}(\vec{r}) \sim 0$. We thus focus on the $\tilde{g}_{a a}(\vec{r})$ in eq.~(\ref{Eq-discus-2}). 

We also obtain the following relations: In the perturbative limit and not so close the resonance, 
\begin{equation}
\tilde{g}_{\alpha \beta} \big| \phi_{\beta} \big|^{2}
\sim 0 
\label{Eq-discus-9}
\end{equation}
because $\phi_{\beta}(\vec{r}) \sim 0$ near vacuum. In the semi-classical limit, 
\begin{equation}
\frac{( h_{\alpha}(\vec{r}) \phi_{\alpha}(\vec{r}) )}{\phi_{\alpha}(\vec{r})}
\sim V_{\alpha}(\vec{r}) + M_{\alpha} 
\label{Eq-discus-7}
\end{equation}
owing to eq.~(\ref{Eq-form-6}) and the zero momenta of BECs, where the zero-point kinetic energies are neglected. 

By substituting eqs.~(\ref{Eq-discus-9}) and (\ref{Eq-discus-7}) into eq.~(\ref{Eq-discus-4}) with eqs.~(\ref{Eq-form-6-2}), (\ref{Eq-discus-5}), and (\ref{Eq-discus-6}), we obtain 
\begin{equation}
\tilde{g}_{a a}(\vec{r}) 
\sim \tilde{g}_{a a}^{(\text{sc})} 
\equiv \frac{4 \pi a_{a a}}{m_{a}} - \frac{2 \lambda^{2}}{\Delta{E}}.
\label{Eq-discus-1}
\end{equation}
If $V_{m}(\vec{r}) \approx 2 V_{a}(\vec{r})$ as assumed in eq.~(\ref{Eq-result-1}), the $\Delta{E}$ in eq.~(\ref{Eq-discus-1}) does not have the spatial dependence. 

The effective interaction $\tilde{g}_{a a}^{(\text{sc})}$ in eq.~(\ref{Eq-discus-1}) agrees with that in the single-channel model, where the resonance changes $a_{a a}$ into $\tilde{a}_{a a}(\Delta{E})$ in eq.~(\ref{Eq-form-10}) and $g_{a a}$ into $\tilde{g}_{a a}^{(\text{sc})}$ in eq.~(\ref{Eq-discus-1}) according to eq.~(\ref{Eq-form-8}). 

Except for the limit, the results in the single-channel model disagree with those in the full model. We label the quantities in the single-channel model with the superscript $(\text{sc})$ to distinguish them from the exact ones. 

In general, the perturbative approach must be valid when the perturbation is small and the nonperturbative state is appropriately selected. The former condition does not matter here because of the regularization in eq.~(\ref{Eq-discus-1}), which corresponds to partial sum of the ladder-type perturbation series; The latter condition must be verified by comparing the perturbative approach with the exact approach. 

\subsection{The role of the molecules \label{Sec-discus-role}}
In the atom dominant regime, the phase separation occurs between the $a$ and $b$ atoms and thus makes the double-domain structure not only in the full system but also in the partial system. 

In the macroscopic analysis for the partial system, the stability condition suggests the phase separation if 
\begin{equation}
\tilde{g}_{a a}(\vec{r}) \tilde{g}_{b b} - \tilde{g}_{a b} \tilde{g}_{b a} 
< 0. 
\label{Eq-discus-role-1}
\end{equation}
It is obtained in the same manner for the full system and becomes 
\begin{equation}
\tilde{g}_{a a}(\vec{r}) 
< g 
\label{Eq-discus-role-2}
\end{equation}
in the situations in the previous section. Eq.~(\ref{Eq-discus-role-2}) is always true and indicates the phase separation because the first term in the right-hand side in eq.~(\ref{Eq-discus-4}) equals to $g$ and the second term is negative in the equilibrium~\footnote{%
The effective scattering length in the equilibrium can not show the drastic sign reversal at the resonance, differently from that in vacuum in eq.~(\ref{Eq-form-10}), because the resonant correlation is always attractive in the equilibrium. 
}. This result is consistent with that for the full system. 

The role of the molecules in the phase separation is thus given in $\tilde{g}_{a a}(\vec{r})$ in the atom dominant regime, where the minority molecules hardly affect the other effective quantities, i.e., $\tilde{\mu}_{\alpha}(\vec{r}) \sim \mu_{\alpha}$, and behave as the resonant intermediary. It is clearly described in eq.~(\ref{Eq-discus-1}) in the single-channel model limit and generally described in eq.~(\ref{Eq-discus-4}) including the medium effect in the equilibrium. The difference between the results in the single-channel and full models is due to the medium effect. 
\subsection{Comparison \label{Sec-discus-compar}}
We here show the difference in the same situation in the previous section: $g = 0.1 \times m^{- 3 / 2} \omega^{- 1 / 2}$, $\kappa = 2$, $\lambda = 0.1 \times m^{- 3 / 4} \omega^{1 / 4}$, and $N_{a t} = N_{b} = 5 \times 10^{4}$. 

\begin{figure}[ht]
\begin{center}
\includegraphics[width=7cm]{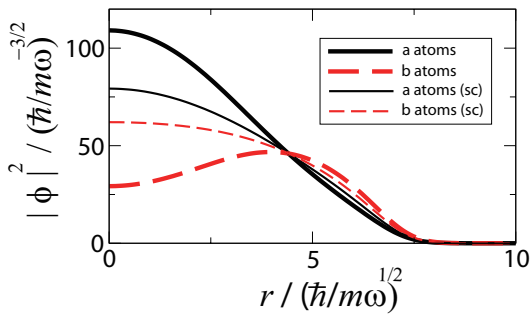}
\end{center}
\caption{(Color online) 
The density profiles when $g = 0.1 \times m^{- 3 / 2} \omega^{- 1 / 2}$, $\kappa = 2$, $\lambda = 0.1 \times m^{- 3 / 4} \omega^{1 / 4}$, $N_{a t} = N_{b} = 5 \times 10^{4}$, and $N_{m} = 5 \times 10^{2}$. The solid and dashed lines denote those of the $a$ and $b$ atoms in the present model; the thin-solid and thin-dashed lines denote those of the $a$ and $b$ atoms in the single-channel model, respectively. 
}
\label{den12}
\end{figure}

In fig.~\ref{den12}, we plot the density profiles in the single-channel and full models for $N_{m} / N_{a t} = 10^{-2}$. The profiles in the full model are same as those in fig.~\ref{den}b. Note that the $\Delta{E}$ in eq.~(\ref{Eq-discus-1}) is determined as a function of $N_{m} / N_{a t}$. 

Although the domain structures are similar in the two models, fig.~\ref{den12} represents obvious difference in the profiles. The difference is mainly caused by the medium effect ignored in eq.~(\ref{Eq-discus-9}) because this case is near the semi-classical limit and not so near the perturbative limit. In fact, if we introduce the effect ignored in eq.~(\ref{Eq-discus-9}) to the single-channel model, the improved results almost agree with those in the full model. 

In order to evaluate fidelities of the order parameters in the single-channel model, we define 
\begin{equation}
F_{\alpha} 
\equiv \frac{1}{N_{\alpha}} \left| \int d{\vec{r}}~ \phi_{\alpha}^{*}(\vec{r}) \phi_{\alpha}^{(\text{sc})}(\vec{r}) \right| 
\label{Eq-discus-10}
\end{equation}
for $\alpha = a$ and $b$, where 
\begin{equation}
N_{\alpha} 
= \int d{\vec{r}}~ \big| \phi_{\alpha}(\vec{r}) \big|^{2} 
= \int d{\vec{r}}~ \big| \phi_{\alpha}^{(\text{sc})}(\vec{r}) \big|^{2}. 
\label{Eq-discus-11}
\end{equation}
The $F_{\alpha}$ in eq.~(\ref{Eq-discus-10}) approaches to one when $\phi_{\alpha}^{(\text{sc})}(\vec{r})$ and $\phi_{\alpha}(\vec{r})$ are similar. 

\begin{figure}[ht]
\begin{center}
\includegraphics[width=7cm]{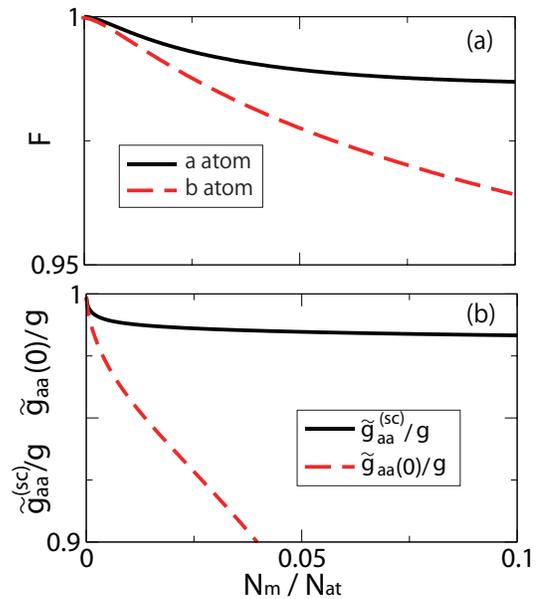}
\end{center}
\caption{(Color online) 
The fidelities (a) and effective coupling constants (b) when $g = 0.1 \times m^{- 3 / 2} \omega^{- 1 / 2}$, $\kappa = 2$, $\lambda = 0.1 \times m^{- 3 / 4} \omega^{1 / 4}$, and $N_{a t} = N_{b} = 5 \times 10^{4}$. The solid and dashed lines denote $F_{a}$ and $F_{b}$ in fig.~\ref{compare}a, respectively, and $\tilde{g}_{a a}^{(\text{sc})} / g$ and $\tilde{g}_{a a}(0) / g$ in fig.~\ref{compare}b. 
}
\label{compare}
\end{figure}

In fig.~\ref{compare}a, we plot the $F_{a}$ and $F_{b}$ in eq.~(\ref{Eq-discus-10}), which gradually decrease as $N_{m}$ increases. Here $F_{b}$ is always smaller than $F_{a}$ because the fidelities reflect the volume factor and the $b$ atoms occupy the outer side. 

In fig.~\ref{compare}b, we plot the $\tilde{g}_{a a}(0)$ in eq.~(\ref{Eq-discus-4}) and $\tilde{g}_{a a}^{(\text{sc})}$ in eq.~(\ref{Eq-discus-1}), where the difference between them gradually increases as $N_{m}$ increases in consistency with fig.~\ref{compare}a. Here $\tilde{g}_{a a}(0)$ is smaller than $\tilde{g}_{a a}^{(\text{sc})}$ because of the medium effect shown in fig.~\ref{den12}. In fact, if we assume eq.~(\ref{Eq-discus-7}), $\phi_{m}(\vec{r}) \sim 0$, and $\tilde{g}_{a a}(\vec{r}) \sim g$, the $\mu_{m}^{\prime}(\vec{r})$ in eq.~(\ref{Eq-discus-4}) becomes 
\begin{equation}
\mu_{m}^{\prime}(\vec{r}) 
\sim - \Delta{E} - \frac{3 \kappa - 13}{8} g \left[ \big| \phi_{a}(\vec{r}) \big|^{2} + \big| \phi_{b}(\vec{r}) \big|^{2} \right], 
\label{Eq-discus-compar-1}
\end{equation}
and then $- \mu_{m}^{\prime}(\vec{r}) < \Delta{E}$ and $\tilde{g}_{a a}(\vec{r}) < \tilde{g}_{a a}^{(\text{sc})}$ for $\kappa = 2$. 

The spatial dependence of $\tilde{g}_{a a}(\vec{r})$ does not play any active roles here~\footnote{%
Here $\tilde{g}_{a a}(\vec{r})$ is a monotonically increasing function for the radial coordinate $r$ and approaches to $g$ at large $r$. 
}. In fact, if we use $\tilde{g}_{a a}(0)$ for the single-channel model instead of $\tilde{g}_{a a}^{(\text{sc})}$ in fig.~\ref{den12}, the improved results mostly reproduce the original profiles and the corresponding fidelities take almost one ($> 0.999$).


\section{Summary and perspective \label{Sec-summary}}

In this paper, we have studied the phase separation of the atomic and molecular BECs with the homonuclear Feshbach resonance by developing the full model formulated in section~\ref{Sec-form}. 

The Feshbach resonance can induce the phase separation, which produces the unsharp domain structures in the trap system as demonstrated in section~\ref{Sec-result}: the double-domain structure in the atom dominant regime and triple-domain structure in the atom-molecule coexistence regime. Here the molecules behave not only as the resonant intermediary but also as another BEC different from the atomic BECs. 

In the atom dominant regime, the role of the molecules is limited to the resonant intermediary and revealed through the effective interactions as discussed in section~\ref{Sec-discus}. Furthermore the effective interactions provide the single-channel model in the perturbative and semi-classical limit. The validity of the single-channel model depends on the medium effect. 

We here comment on the recent experiments of the artificial phase separation of BECs in the Feshbach resonance technique~\cite{papps,tojo}. The experiments are performed in the atom dominant regime, and the phase separation refers to that of the atomic BECs in analogy with the double-domain structure in the present work. The triple-domain structure has not been realized in actual experiments. We are expecting further development of the experimental study in this field. 

The actual experiments are usually performed in anisotropic traps, not in spherical traps as in the present work. The asymmetry of the traps can induce anisotropic domain structures as shown in the recent experiments~\cite{papps,tojo}. In principle, the anisotropic case can be described in the same framework in this paper by adopting the anisotropic traps as $V_{\alpha}(\vec{r})$ in eq.~(\ref{Eq-form-6}). 

We also comment on the previous study on the quasi-chemical equilibrium theory for the BECs~\cite{TN}. The concept and details of the work are quite different from those of the present work: the macroscopic effective theory vs. the microscopic model, the heteronuclear molecules vs. the homonuclear molecules, the negligible correlation vs. the resonant correlation, and so on. However two significant analogies are seen in the results in both works. First the atom-molecule phase separation appears as the coexistence phases in the previous work and triple-domain structure in the present work. It is due to the interparticle interactions and reflects the repulsion between the atomic and molecular BECs. Second the interaction energies can contribute to the results as the renormalized energy detuning in the previous work and medium effect in the present work. 

In the macroscopic analysis, some additional works have recently been published. In ref.~\cite{ammix2}, the quasi-chemical equilibrium theory is developed to the case near the heteronuclear Feshbach resonance with the resonant correlation. In ref.~\cite{ammix}, the homonuclear molecule case is studied for the one-component atoms. 

We here emphasize that the macroscopic analysis can not give any information on the actual domain structures. In fact, e.g., the occurrence of the coexistence phases in the quasi-chemical equilibrium theory merely indicates the possible existence of the triple-domain structure and does not determine the appearance and profile, which are determined in the microscopic analysis as demonstrated in this paper. 

Finally we notice the application limit of the present approach in terms of the validity of the Bogoliubov's classical field approximation, which must be valid in the BEC dominant regime and not be applied to the strongly correlated Bose gases. If the resonant correlation is quite strong and the situation is near the molecule dominant regime, the atoms may behave as the strongly correlated gas in analogy with the nondegenerate fermionic atoms in the BEC-BCS crossover. This situation is beyond the purpose of this paper and should be studied in another paper. 

\acknowledgments

This work was supported by the MEXT program "Support Program for Improving Graduate School Education" and KAKENHI (22540414). 

\appendix

\section{Feshbach resonance \label{App-fr}}
Let us consider elastic scattering between two atoms in vacuum with a Feshbach resonance. The Schr\"{o}dinger equation can be written as coupled channel equations, 
\begin{equation}
\left( E - P H P \right) \big| \Psi_{E}^{P} \big> 
= \left( P H Q \right) \big| \Psi_{E}^{Q} \big>, 
\label{Eq-fr-1}
\end{equation}
\begin{equation}
\left( E - Q H Q \right) \big| \Psi_{E}^{Q} \big> 
= \left( Q H P \right) \big| \Psi_{E}^{P} \big>, 
\label{Eq-fr-2}
\end{equation}
for $\big| \Psi_{E}^{P} \big> \equiv P \big| \Psi_{E} \big>$ and $\big| \Psi_{E}^{Q} \big> \equiv Q \big| \Psi_{E} \big>$ with the projection operators $P$ ($= P^{\dagger} = P^{2}$) for the open-channel and $Q$ ($= Q^{\dagger} = Q^{2}$) for the closed-channel obeying $P + Q = 1$ and $P Q = Q P = 0$. 

The open-channel Hamiltonian $P H P \equiv H_{\text{o}}$ in eq.~(\ref{Eq-fr-1}) is given as 
\begin{equation}
H_{\text{o}} 
= K_{\text{o}} + V_{\text{o}} 
\label{Eq-fr-3}
\end{equation}
with the kinetic part $K_{\text{o}}$ of the relative motion in the center-of-mass system and the bare interaction part $V_{\text{o}}$ denoted by the contact-type interaction, 
\begin{equation}
\mathfrak{N} \big< \phi_{\vec{q}} \big| V_{\text{o}} \big| \phi_{\vec{p}} \big> 
= g, 
\label{Eq-fr-4}
\end{equation}
where we introduce the arbitrary normalization factor $\mathfrak{N}$ and energy-momentum eigen-state, 
\begin{equation}
K_{\text{o}} \big| \phi_{\vec{p}} \big> 
= \frac{p^{2}}{2 \mathfrak{m}} \big| \phi_{\vec{p}} \big> 
= E \big| \phi_{\vec{p}} \big>, 
\label{Eq-fr-5}
\end{equation}
with the reduced mass $\mathfrak{m}$. 

In the regularization in eq.~(\ref{Eq-form-8}), we renormalize  
\begin{equation}
g 
\to \lim_{E \to 0} \mathfrak{N} \int \frac{d{\Omega_{\vec{p}}}}{4 \pi} \big< \phi_{\vec{p}} \big| V_{\text{o}} \big| \psi_{E} \big> 
= \frac{2 \pi a_{0}}{\mathfrak{m}} 
\label{Eq-fr-6}
\end{equation}
with the solid angle $\Omega_{\vec{p}}$ and two-body scattering wave-function, 
\begin{equation}
\big| \psi_{E} \big> 
=  \left( 1 + \frac{1}{E - H_{\text{o}} + i \epsilon} V_{\text{o}} \right) \big| \phi_{\vec{p}} \big>, 
\label{Eq-fr-7}
\end{equation}
where the bare $s$-wave scattering length $a_{0}$ is related to $g$ as 
\begin{equation}
a_{0} 
=  \frac{g}{4 \pi} \left( 1 + g \Lambda \right)^{-1} 
\label{Eq-fr-8}
\end{equation}
with the divergent integral  
\begin{equation}
\Lambda 
\equiv \int_{0}^{\infty} \frac{d{p}}{2 \pi^{2}}. 
\label{Eq-fr-9}
\end{equation}
The closed-channel Hamiltonian $Q H Q \equiv H_{\text{c}}$ in eq.~(\ref{Eq-fr-2}) gives energy eigen-states in the $Q$-space, 
\begin{equation}
H_{\text{c}} \big| \chi_{n} \big> 
= \varepsilon_{n} \big| \chi_{n} \big>, 
\label{Eq-fr-10}
\end{equation}
including the resonant molecular state $\big| \chi_{M} \big>$ and energy detuning, $\varepsilon_{M} = \Delta{E}$, for $n = M$. 

According to eqs.~(\ref{Eq-fr-1}) and (\ref{Eq-fr-2}), we obtain 
\begin{equation}
\left( E - H_{\text{o}} \right) \big| \Psi_{E}^{P} \big> 
= V_{E} \big| \Psi_{E}^{P} \big> 
\label{Eq-fr-11}
\end{equation}
with the effective interaction 
\begin{equation}
V_{E} 
\equiv \left( P H Q \right) \frac{1}{E - H_{\text{c}} + i \epsilon} \left( Q H P \right), 
\label{Eq-fr-12}
\end{equation}
where we assume 
\begin{equation}
V_{E} 
\approx \left( P H Q \right) \frac{\big| \chi_{M} \big> \big< \chi_{M} \big|}{E - \Delta{E} + i \epsilon} \left( Q H P \right) 
\label{Eq-fr-13}
\end{equation}
owing to the complete set in the $Q$-space, 
\begin{equation}
\sum_{n} \big| \chi_{n} \big> \big< \chi_{n} \big| 
= Q, 
\label{Eq-fr-14}
\end{equation}
and the resonance, $\big| E - \varepsilon_{n \ne M} \big| \gg \big| E - \Delta{E} \big|$. 

The resonance part is denoted by the contact-type correlation, 
\begin{equation}
\sqrt{\mathfrak{N}} \big< \chi_{M} \big| Q H P \big| \phi_{\vec{p}} \big> 
= \sqrt{2} \lambda, 
\label{Eq-fr-15}
\end{equation}
where the coefficient $\sqrt{2}$ is due to symmetrization for the identical atoms. Note that the exchange part, $\big| \phi_{- \vec{p}} \big> / \sqrt{2}$, does not contribute to the molecule formation in eq.~(\ref{Eq-fr-15}). 

In the regularization in eq.~(\ref{Eq-form-10}), we renormalize  
\begin{equation}
\lambda 
\to \lim_{E \to 0} \frac{1}{\sqrt{2}} \frac{\sqrt{\mathfrak{N}}}{\mathfrak{C}_{E}} \big< \chi_{M} \big| Q H P \big| \Psi_{E}^{P} \big> 
\equiv \tilde{\lambda}, 
\label{Eq-fr-16}
\end{equation}
where the two-body wave-function in the open-channel is written as 
\begin{equation}
\big| \Psi_{E}^{P} \big> 
= \mathfrak{C}_{E} \left[ 1 + \frac{1}{E - H_{\text{o}} - V_{E} + i \epsilon} ( V_{\text{o}} + V_{E} ) \right] \big| \phi_{\vec{p}} \big>, 
\label{Eq-fr-17}
\end{equation}
with the dissipation factor $\mathfrak{C}_{E}$ into the molecule according to eq.~(\ref{Eq-fr-11}). The $\lambda$ and $\tilde{\lambda}$ in eq.~(\ref{Eq-fr-16}) obey 
\begin{equation}
\tilde{\lambda} 
= \lambda \left( 1 + \tilde{g} \Lambda \right)^{-1} 
\label{Eq-fr-18}
\end{equation}
with the effective coupling constant 
\begin{equation}
\tilde{g} 
\equiv g - \frac{2 \lambda^{2}}{\Delta{E}}. 
\label{Eq-fr-19}
\end{equation}

The $s$-wave scattering length $\tilde{a}_{0}$ in the open-channel is written as
\begin{equation}
\tilde{a}_{0} 
=  \frac{\tilde{g}}{4 \pi} \left( 1 + \tilde{g} \Lambda \right)^{-1}. 
\label{Eq-fr-20}
\end{equation}
According to eqs.~(\ref{Eq-fr-8}) and (\ref{Eq-fr-18})-(\ref{Eq-fr-20}), we obtain 
\begin{equation}
\tilde{a}_{0} 
= a_{0} - \frac{1}{2 \pi} \frac{\xi^{2}}{\Delta{E} - \delta} 
\label{Eq-fr-21}
\end{equation}
and 
\begin{equation}
\tilde{\lambda} 
= \xi \left( 1 - \frac{\delta}{\Delta{E}} \right)^{-1} 
\label{Eq-fr-22}
\end{equation}
with the resonance shift 
\begin{equation}
\delta 
\equiv \frac{2 \lambda^{2} \Lambda}{1 - 4 \pi a_{0} \Lambda} 
\label{Eq-fr-23}
\end{equation}
and width 
\begin{equation}
\xi 
\equiv \frac{\delta}{\lambda \Lambda}, 
\label{Eq-fr-24}
\end{equation}
which are independent of $\Delta{E}$. In the second-order perturbation of $\lambda$, we obtain $\big| \delta / \Delta{E} \big| \ll 1$ and eq.~(\ref{Eq-form-10}). 
\section{Stability conditions \label{App-psc}}
The energy density $\mathcal{E}[ n_{a}, n_{b}, n_{m} ]$ for the uniform BECs, $\big< H \big> = \int d{\vec{r}}~ \mathcal{E}$, is given as 
\begin{equation}
\mathcal{E} 
= \left[ \sum_{\alpha = a, b, m} \sum_{\beta = a, b, m} \frac{g_{\alpha \beta}}{2} n_{\alpha} n_{\beta} \right] - 2 | \lambda | n_{a} n_{m}^{1 / 2} 
\label{Eq-psc-1}
\end{equation}
with the $g_{\alpha \beta}$ ($= g_{\beta \alpha}$) in eqs.~(\ref{Eq-result-0}), (\ref{Eq-result-3}), and (\ref{Eq-result-4}). 

The phase ($a$, $b$, $m$) is unstable because the Hessian 
\begin{equation}
\mathcal{H}^{(a b m)} 
\equiv \det{\left[ \frac{\partial^{2} \mathcal{E}[ n_{a}, n_{b}, n_{m} ]}{\partial{n_{\alpha}} \partial{n_{\beta}}} \right]} 
\label{Eq-psc-2}
\end{equation}
for $\alpha, \beta = a, b, m$ is negative: 
\begin{equation}
\mathcal{H}^{(a b m)} 
= - \frac{g \lambda^{2}}{n_{m}} 
< 0. 
\label{Eq-psc-3}
\end{equation}

The phase ($b$, $m$) is unstable because the Hessian 
\begin{equation}
\mathcal{H}^{(b m)} 
\equiv \det{\left[ \frac{\partial^{2} \mathcal{E}[ n_{a} = 0, n_{b}, n_{m} ]}{\partial{n_{\alpha}} \partial{n_{\beta}}} \right]} 
\label{Eq-psc-4}
\end{equation}
for $\alpha, \beta = b, m$ is negative: 
\begin{equation}
\mathcal{H}^{(b m)} 
= - \frac{9}{64} g^{2} \left[ \left( \kappa - \frac{7}{9} \right)^{2} + \frac{32}{81} \right] 
< 0. 
\label{Eq-psc-5}
\end{equation}

The phases ($a$, $b$) and ($a$) are unstable because of the asymptotic behaviors near $n_{m} = 0$. 

The phase ($a$, $m$) is stable or meta-stable when the Hessian 
\begin{equation}
\mathcal{H}^{(a m)} 
\equiv \det{\left[ \frac{\partial^{2} \mathcal{E}[ n_{a}, n_{b} = 0, n_{m} ]}{\partial{n_{\alpha}} \partial{n_{\beta}}} \right]} 
\label{Eq-psc-6}
\end{equation}
for $\alpha, \beta = a, m$ is positive: 
\begin{equation}
\mathcal{H}^{(a m)} 
= \mathcal{H}^{(b m)} + \frac{g \big| \lambda \big|}{2 n_{m}^{1 / 2}} \left[ 
\frac{n_{a}}{n_{m}} 
+ \frac{3 ( 1 + \kappa )}{2} \right] 
- \frac{\lambda^{2}}{n_{m}} 
\geq 0. 
\label{Eq-psc-7}
\end{equation}

The phase ($b$) is always stable or meta-stable. 

The phase ($m$) is stable or meta-stable when 
\begin{equation}
n_{m} 
> \left[ \frac{16}{3} \frac{\lambda}{( 1 + \kappa ) g} \right]^{2} 
\label{Eq-psc-8}
\end{equation}
because of the asymptotic behavior near $n_{a} = n_{b} = 0$.

\end{document}